\documentclass[11pt,letterpaper]{article}
\pdfoutput=1
\usepackage{amsmath,amsfonts,amssymb,latexsym}
\usepackage{graphicx}
\usepackage[raggedright]{subfigure}

\newcommand{\subG}{\ensuremath{\mathcal{S}(G)}}
\newcommand{\objG}{\ensuremath{\mathcal{O}(G)}}
\newcommand{\predG}{\ensuremath{\mathcal{P}(G)}}
\newcommand{\allG}{\ensuremath{\mathcal{A}(G)}}
\newcommand{\atoms}{\ensuremath{\mathcal{A}}}

\newtheorem{thm}{Theorem}
\newtheorem{ex}[thm]{Example}

\begin{document}

\title{A Role-Free Approach to Indexing Large RDF Data Sets in Secondary Memory
for Efficient SPARQL Evaluation}
\author{George H.\ L.\ Fletcher and Peter W.\ Beck\\
       School of Engineering and Computer Science\\
       Washington State University, Vancouver, USA\\
       {\normalsize $\{${\tt fletcher}, {\tt pwbeck}$\}${\tt @wsu.edu}}}
\date{}
\maketitle

\section{Introduction}

Massive RDF data sets are becoming commonplace.  RDF data is typically generated
in social semantic domains (such as personal information management
\cite{Aberer06,KargerBHQS05,schraefel07}) wherein a fixed schema is often not
available a priori.  We propose a simple {\em Three-way Triple Tree} (TripleT)
secondary-memory indexing technique to facilitate efficient SPARQL query
evaluation on such data sets.  The novelty of TripleT is that (1) the index is
built over the atoms occurring in the data set, rather than at a coarser
granularity, such as whole triples occurring in the data set; and (2) the atoms
are indexed regardless of the roles (i.e., subjects, predicates, or objects)
they play in the triples of the data set.  We show through extensive empirical
evaluation that TripleT exhibits multiple orders of magnitude improvement over
the state of the art on RDF indexing, in terms of both storage and query
processing costs.

\paragraph{Preliminary Notions.}  We assume familiarity with the RDF and SPARQL standards
\cite{FurcheLBPG06,KlyneC04,PrudS08}, the B+tree data structure \cite{Comer1979,RG}, and the
basics of conjunctive query processing \cite{ChandraM77,RG,SelingerACLP79}.  Let
\atoms\ be an enumerable set of atoms (e.g., Unicode strings). A {\em triple} is 
an element of $\atoms\times\atoms\times\atoms$.
An RDF {\em graph} is a finite set of triples.
For graph $G$, let 
\begin{eqnarray*}
\subG  &=& \{s \mid (s, p, o)\in G\} \\
\predG &=& \{p \mid (s, p, o)\in G\} \\
\objG  &=& \{o \mid (s, p, o)\in G\} \\
\allG &=& \subG \cup \predG \cup \objG.
\end{eqnarray*}
The atoms appearing in \subG\ are called the {\em subjects} of $G$;
the atoms appearing in \predG\ are called the {\em predicates} of $G$; and, 
the atoms appearing in \objG\ are called the {\em objects} of $G$.


\section{The Problem}

The problem we consider in this paper is how to index a graph $G$ to
support efficient evaluation of {\em basic graph patterns} (BGP) over $G$. BGPs,
which are conjunctions of {\em simple access patterns} (SAP), form the heart of
all SPARQL queries. 

\begin{figure}[!t]
\begin{center}
{\tt 
{\footnotesize
\begin{tabular}[b]{l@{~~}l@{~~}l}
$\big\{\langle$Yamada, & authored, & doc1$\rangle$, \\
$~\langle$Yamada, & knows, & McShea$\rangle$, \\
$~\langle$knows, & is a kind of, & social action$\rangle$, \\
$~\langle$Herzog, & authored, & doc2$\rangle$, \\
$~\langle$Herzog, & authored, & doc3$\rangle$, \\
$~\langle$McShea, & performed, & doc3$\rangle$, \\
$~\langle$McShea, & past action, & authored$\rangle$, \\
$~\langle$doc1, & type, & PDF$\rangle$, \\
$~\langle$doc1, & rating, & 4/5$\rangle$, \\
$~\langle$doc2, & type, & MP3$\rangle,$ \\
$~\langle$doc3, & type, & MP3$\rangle$, \\ 
$~\langle$doc3, & created on, & 26.10.08$\rangle\big\}$\\
~&~&~
\end{tabular}
}}
\end{center}
\caption{A triple graph.}\label{fig:main}
\end{figure}

\begin{ex}\label{ex:sparql}

Consider the query ``What are the dates and types of documents on which McShea
was a performer?'' over the triple store given in Figure \ref{fig:main}.  
In SPARQL, where variables are identified by a leading {\tt ?},
this query can be formulated as follows:

\begin{center}
\begin{minipage}{.65\textwidth}
{\footnotesize
\begin{verbatim}
SELECT ?date ?type
WHERE { McShea performed  ?doc  . 
        ?doc   created_on ?date .
        ?doc   type       ?type  } 
\end{verbatim}
}
\end{minipage}
\end{center}

\noindent The {\tt WHERE} clause of a SPARQL query specifies a BGP, 
which in this case consists
of the conjunction of the following three SAPs:
$$(\mathtt{McShea}, \mathtt{performed}, \textnormal{?doc}),
(\textnormal{?doc}, \mathtt{created~on}, \textnormal{?date}),
(\textnormal{?doc}, \mathtt{type}, \textnormal{?type}).$$
Conceptually, the evaluation of a BGP on a graph $G$ consists of 
finding all variable bindings 
such that each of the BGP's constituent SAPs simultaneously holds in $G$.
In our example, there is only one set of valid
variable bindings:
\begin{center}
\begin{tabular}{c@{~~~~~~}c@{~~~~~~}c}
$\textnormal{?doc}$ & $\textnormal{?date}$ & $\textnormal{?type}$  \\
\hline
\tt{doc1}& \tt{26.10.08}& \tt{MP3}
\end{tabular}
\end{center}
The {\tt SELECT} clause indicates that only the bindings for
\textup{?date} and \textup{?type} are returned in the query result.
\end{ex}

The reader will recognize that BGPs are essentially conjunctive 
queries evaluated over a single ternary relation 
\cite{ChandraM77,Fletcher08,GutierrezHM04,SelingerACLP79,StockerSBKR08}.
Joins between the SAPs of a BGP are induced by the co-occurrence of variables
and atoms.  There are six native BGP join types: subject-subject,
subject-predicate, subject-object, predicate-predicate, predicate-object, and
object-object joins.  In Example \ref{ex:sparql}, there is a subject-object join between the
first SAP and both the second and third SAPs, due to the
co-occurrence of variable ?doc.  Furthermore, there is a subject-subject join between
the second and third SAPs.

We specifically focus on the problem of designing {\em native} RDF index data structures to
accelerate BGP evaluation.  By native, we mean data structures which support the
full range of BGP join patterns.

\section{The Solution}

Let $G$ be a fixed RDF graph.  In what follows,  we use the B+tree
secondary-memory data structure \cite{Comer1979} to implement the various
indexing techniques considered.  However, any of a variety of appropriate
secondary-memory data structures (e.g., linear hashing \cite{RG}) could also
be also have been used.

\begin{figure}[!t]
\begin{center}
\subfigure[MAP]{\label{fig:map}
\begin{tabular}[b]{c}
\includegraphics[width=0.29\textwidth]{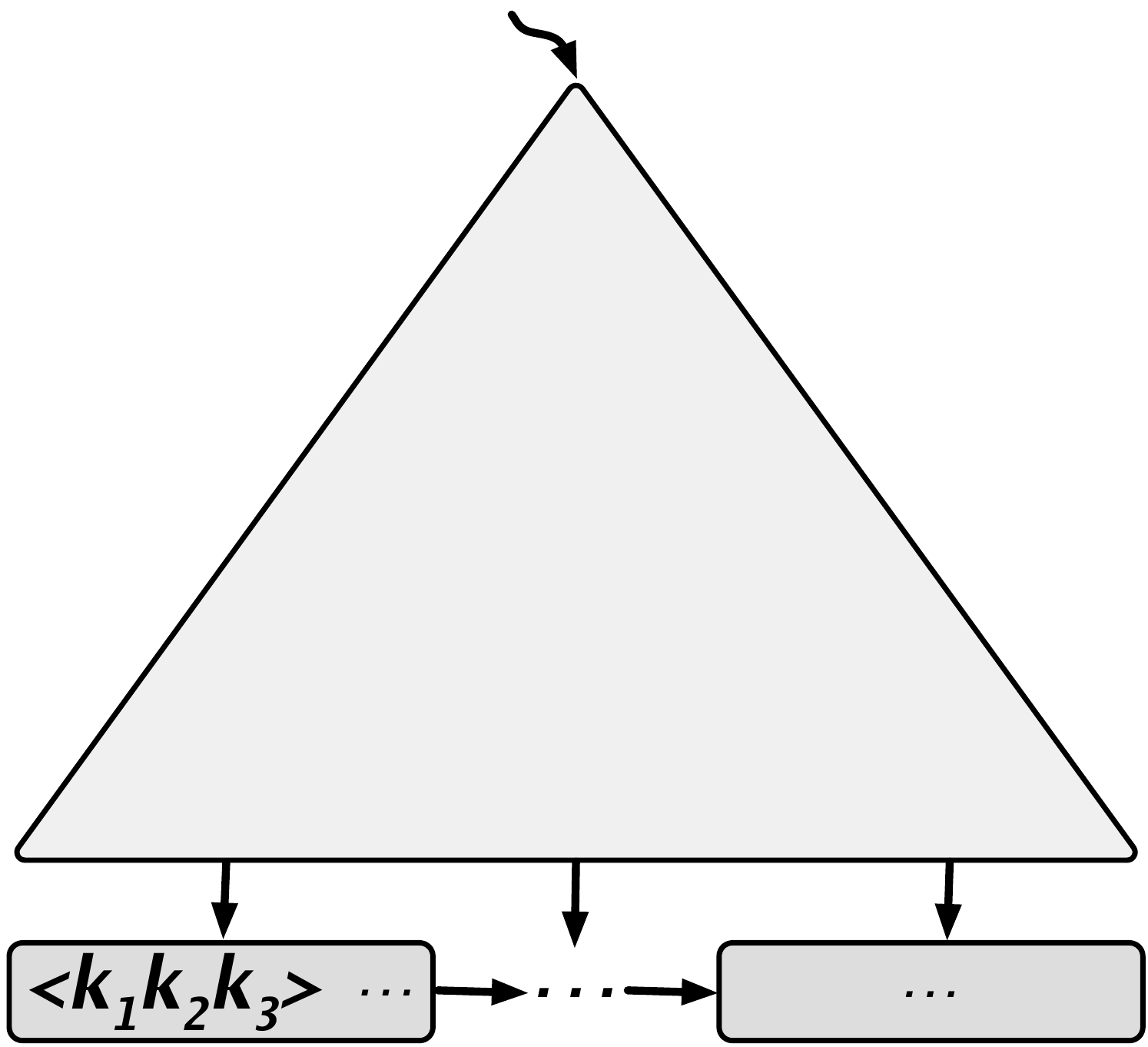}\\
~\\
~
\end{tabular}
}
\hspace{\stretch{1}}
\subfigure[HexTree]{\label{fig:hex}
\includegraphics[width=0.29\textwidth]{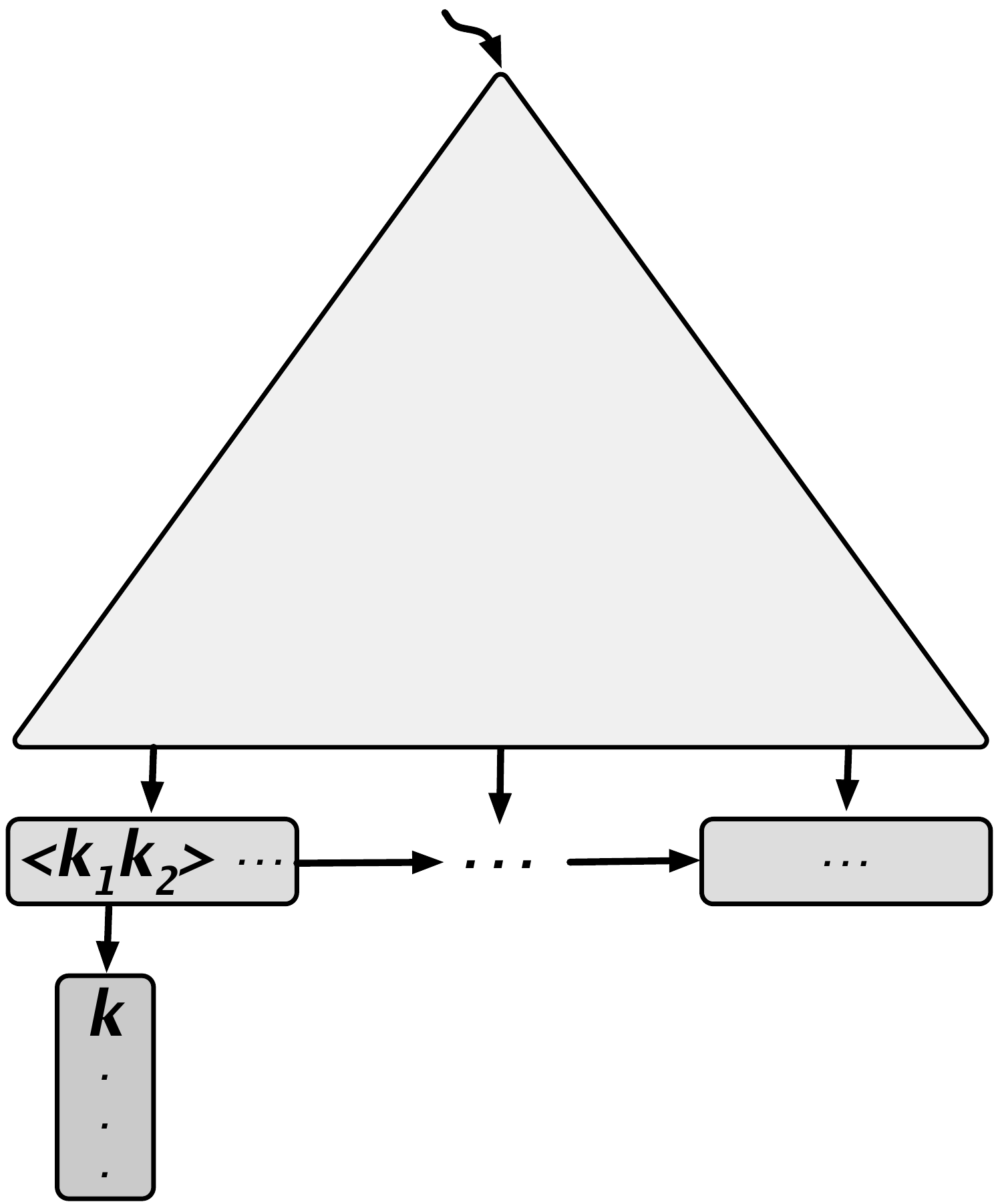}}
\hspace{\stretch{1}}
\subfigure[TripleT]{\label{fig:triplet}
\includegraphics[width=0.29\textwidth]{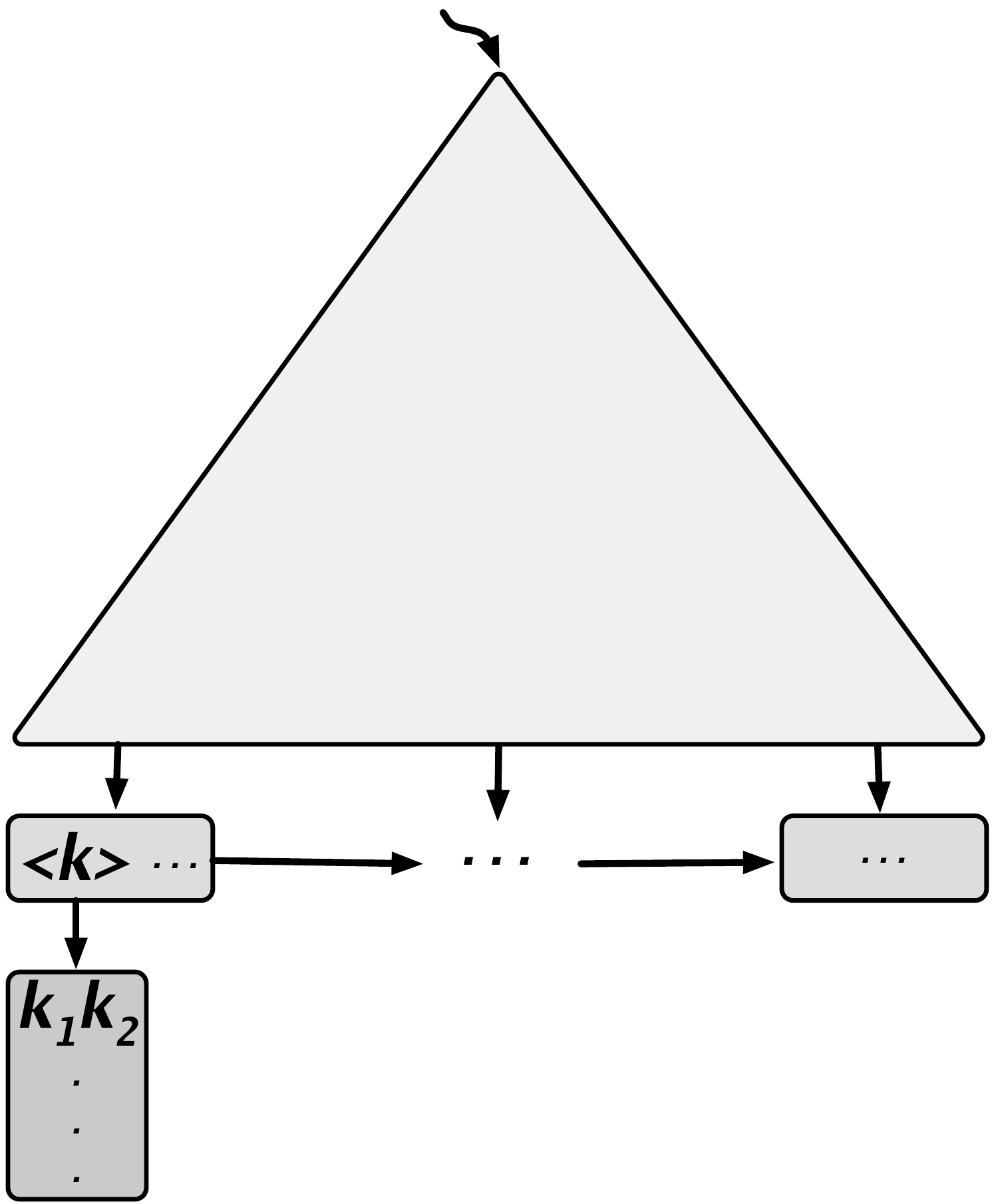}}
\caption{Varieties of Triple Trees.}
\end{center}
\end{figure}
\subsection{State of the Art}

To the best of our knowledge, the two major competitive proposals for native RDF
indexing are multiple access patterns (MAP) and HexTree.

\begin{itemize}

    \item {\bf MAP.} In this approach, all three positions of triples are
    indexed: subjects (S), predicates (P), and objects (O), for some permutation
    of S, P, and O.  MAP requires up to six separate indexes, corresponding to
    the six possible orderings of roles: SPO, SOP, PSO, POS, OSP, OPS.  For
    example, for each $(s, p, o)\in G$, it is the case that $o\#p\#s$ is a key
    in the OPS index on $G$; see Figure \ref{fig:map}.\footnote{Where ``$\#$'' is
    some reserved separator symbol.} A BGP join evaluation requires two or more
    look-ups, potentially in different trees, followed by merge-joins.  Major
    systems employing this technique include Virtuoso, YARS, RDF-3X, Kowari, and
    System-$\Pi$ \cite{erling08,HarthUHD07,NeumannW08,WoodGA05,WuLW08}.  In the
    present investigation we use the B+tree data structure for each of the MAP
    indexes (Figure \ref{fig:map}).

    \item {\bf HexTree.} Recently in the Hexstore system, Weiss et al.\
    \cite{weissKB08} have proposed indexing two roles at a time.  This approach
    requires up to six separate indexes corresponding to the six possible
    orderings of roles: SO, OS, SP, PS, OP, PO.  Payloads are shared between
    indexes with symmetric orderings.  For example, for each $(s, p, o)\in G$, it
    is the case that $s\#p$ is a key in the SP index on $G$, $p\#s$ is a key in
    the PS index on $G$, and both of these keys point to a payload of
    $\{o\in\objG\mid (s, p, o)\in G\}$; see Figure \ref{fig:hex}.  As with MAP,
    join evaluation requires two or more look-ups, potentially in different
    trees, followed by merge-joins.  Hexstore has only been proposed and
    evaluated as a main-memory data structure \cite{weissKB08}.  We propose
    HexTree as an effective secondary-memory realization of the Hexstore
    proposal using the B+tree data structure (Figure \ref{fig:hex}).

\end{itemize}

Note that techniques have also been developed for indexing heuristically-selected classes
of larger graph patterns, e.g., \cite{UdreaPS07}. Such techniques, however, do
not support processing of the full range of native BGP join patterns.

\subsection{Our Proposal}

\begin{figure}[!t]
\centering
\includegraphics[width=0.2\textwidth]{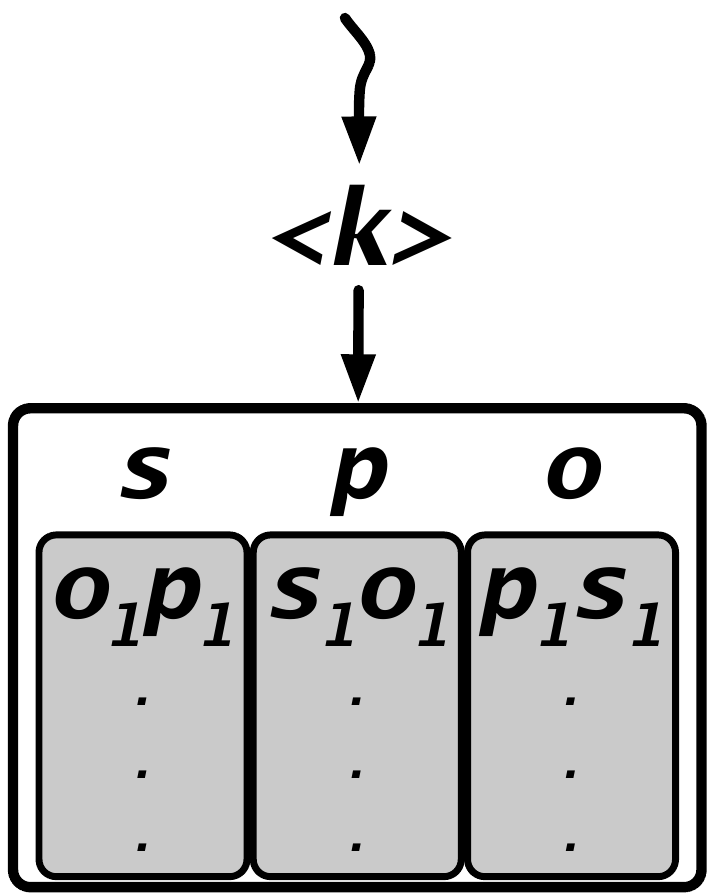}
\caption{TripleT payload for atom $k$.}\label{fig:payload}
\end{figure}

We propose indexing the key-space \allG,
regardless of the particular roles the atoms of \allG\ play in the triples of $G$.
For a key $k$, the payload is all triples of $G$ in which atom $k$ occurs (see Figure
\ref{fig:triplet}).  In particular, the payload for $k$ consists of three
``buckets'':  
one for all pairs $(p, o)$ where $(k, p, o)\in G$, 
one for all pairs $(s, o)$ where $(s, k, o)\in G$, and
one for all pairs $(s, p)$ where $(s, p, k)\in G$, (see Figure \ref{fig:payload}). 
In other words, there is one bucket apiece for all those triples where $k$
occurs as a subject, for all those triples where $k$ occurs as a predicate, and
for all those triples where $k$ appears as an object.  For example, on the graph
of Figure \ref{fig:main},  the payload for {\tt doc1} would consist of an object
bucket $\langle(\mathtt{Yamada}, \mathtt{authored})\rangle$, a subject bucket $\langle
(\mathtt{4/5}, \mathtt{rating}), (\mathtt{PDF}, \mathtt{type}) \rangle$, and a
predicate bucket $\langle\rangle$.\footnote{To facilitate query processing, note that we
keep the pairs in each of the buckets sorted.  By default, the subject bucket is
sorted in OP order, the predicate bucket in SO order, and the object bucket in
SP order.}  
TripleT requires just one index, while efficiently supporting all join patterns
native to SPARQL.  For example, a subject-object join induced by the
co-occurrence of an atom $k$ can be evaluated by a single look-up on $k$
followed by a merge-join between the subject and object buckets of $k$'s
payload.  A join induced by the co-occurrence of a variable is implemented as
multiple look-ups followed by merge-joins, as with MAP and HexTree.  However,
since the keys in TripleT are 1/3 the length of those in MAP and 1/2 those in
HexTree, there is a significant increase in the branching factor of the TripleT
B+tree, which leads to a significant reduction in cost for these look-ups.

TripleT does not favor any particular join types, supporting the full range of
join patterns native to RDF data.  The recently proposed
``vertical-partitioning'' approach \cite{AbadiMMH07} can be viewed as a special
restricted case of TripleT where (1) only the atoms of \predG\ are indexed and
(2) only the predicate payload bucket for each key is maintained.  In this
sense, vertical-partitioning is not a fully native RDF indexing technique;
indeed, recent research has demonstrated practical limitations of this
approach \cite{Sidirourgos08,SchmidtHKLP08,weissKB08}.  This research has also
demonstrated similar limitations of the related ``property table'' RDF storage
techniques \cite{GarciaM06,SintekK06,TheoharisCK05,wilkinson06}.

\section{Empirical Evaluation}\label{sec:eval}

We implemented all three approaches using 8K blocks and 32-bit references, in
virtual memory, using Python 2.5.2.  All experiments were executed
on a pair of 2.66 GHz dual-core Intel Xeon processors with 16 GB RAM running Mac OS X 10.4.11.
Each experiment was performed using (1) simple synthetic data; (2) the DBPedia
RDF data set; and, (3) the Uniprot RDF data set.  Further details of these data
sets are provided in the Appendix.

As mentioned above, in TripleT we only materialized the OP, SO, and SP sort
orderings for the subject, predicate, and object payload buckets,
respectively.\footnote{If necessary, each of the two possible sort orderings for
each of the three TripleT buckets could be materialized.  In this case, we would
of course still need just one B+tree to index payloads.}  Consequently, we only built the
corresponding 
SOP, PSO, and OSP trees for MAP and the SO, PS, and OS trees for HexTree.  In
all of our experiments, the TripleT payloads occupied on average only one disk
block.  Hence, if a symmetric sort ordering was necessary for a merge join
(e.g., if the PO ordering was necessary for the subject bucket while using
TripleT or if the SPO ordering was necessary while doing a lookup in MAP), the
sort was performed in main-memory without penalty.

\subsection{Index size}

\begin{figure}[!t]
\begin{center}
\subfigure[Synthetic]{\label{fig:sizesynth}
\includegraphics[width=0.33\textwidth]{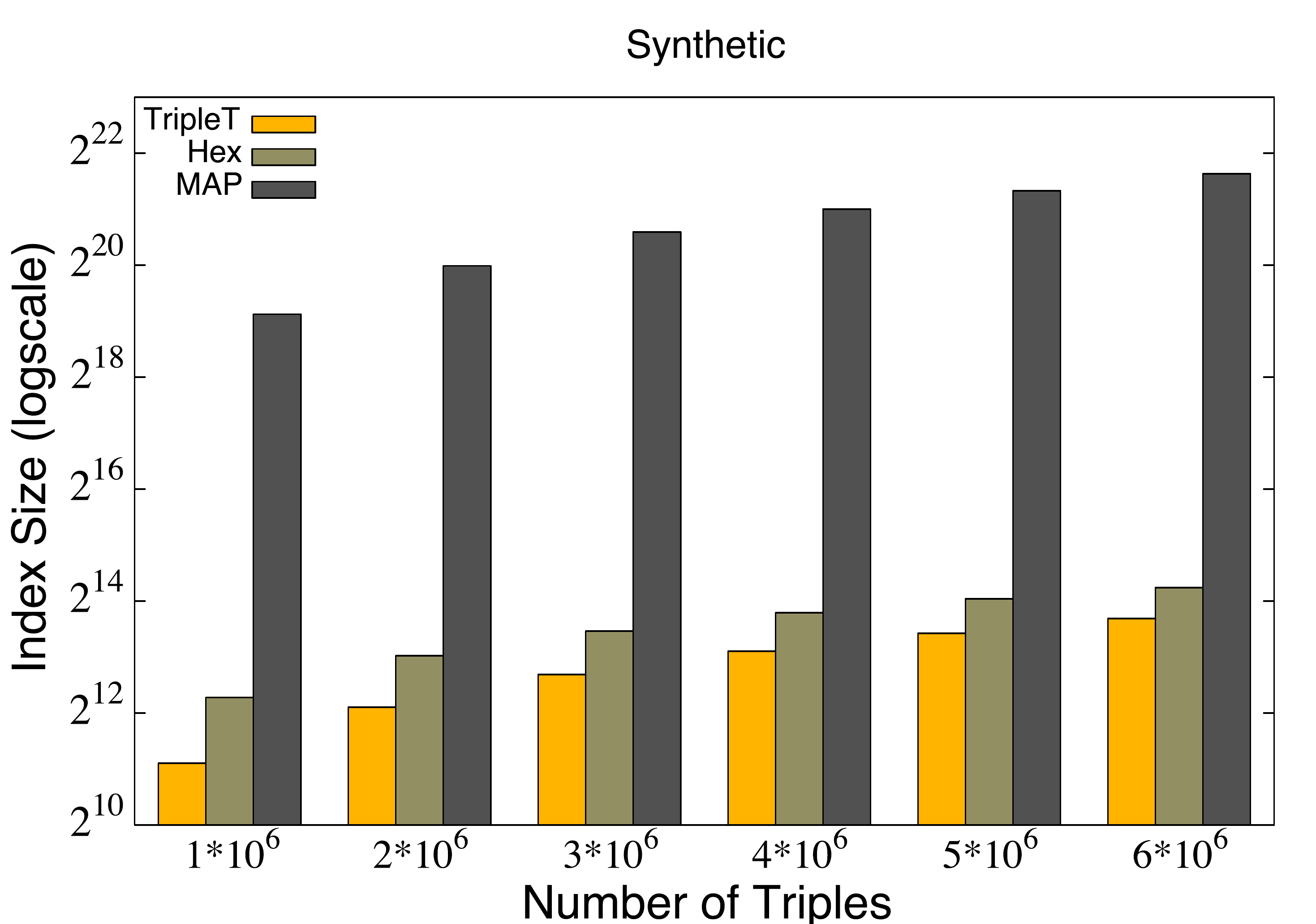}}
\hspace{-.4cm}
\subfigure[DBPedia]{\label{fig:sizedb}
\includegraphics[width=0.33\textwidth]{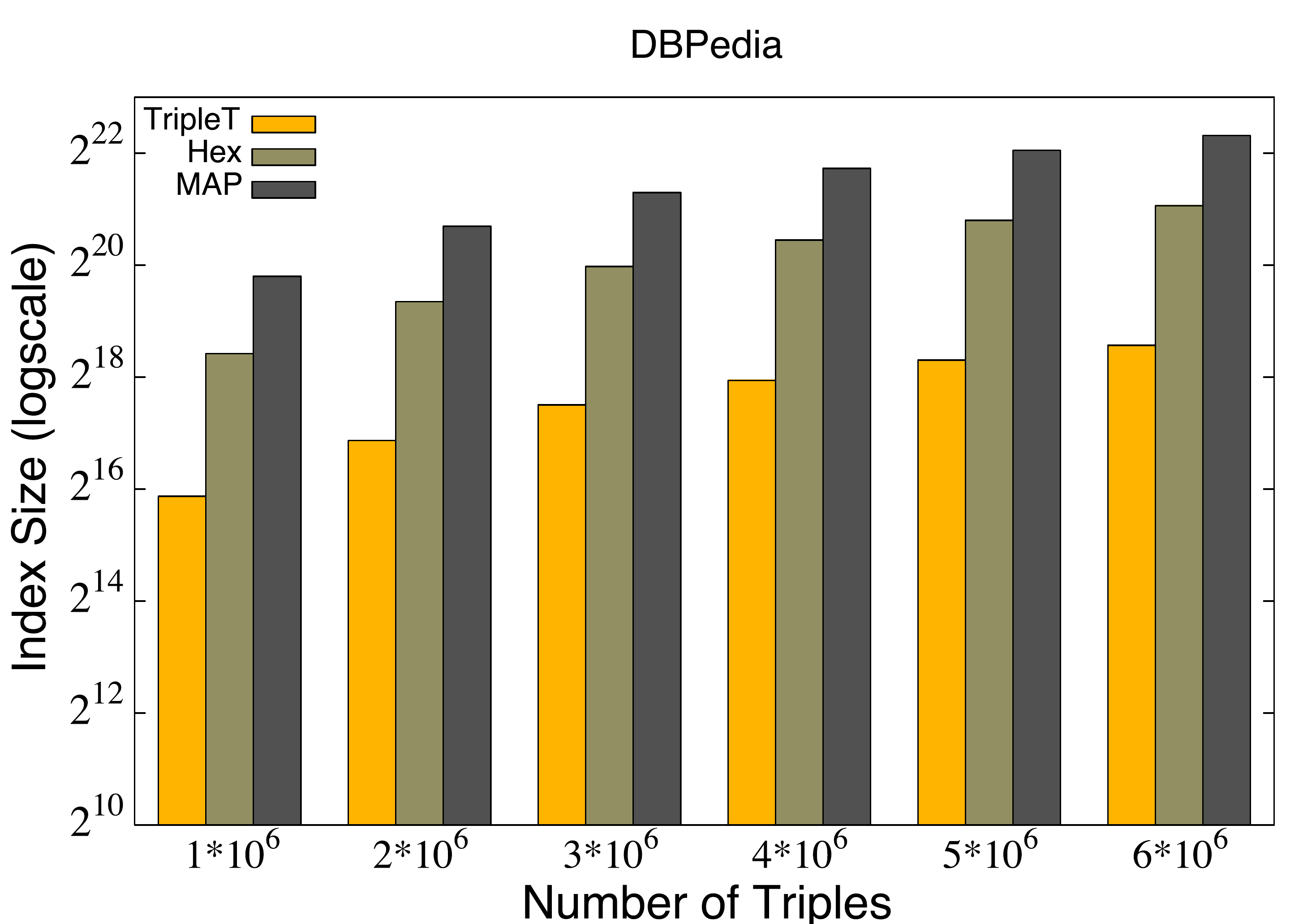}}
\hspace{-.4cm}
\subfigure[Uniprot]{\label{fig:sizeuni}
\includegraphics[width=0.33\textwidth]{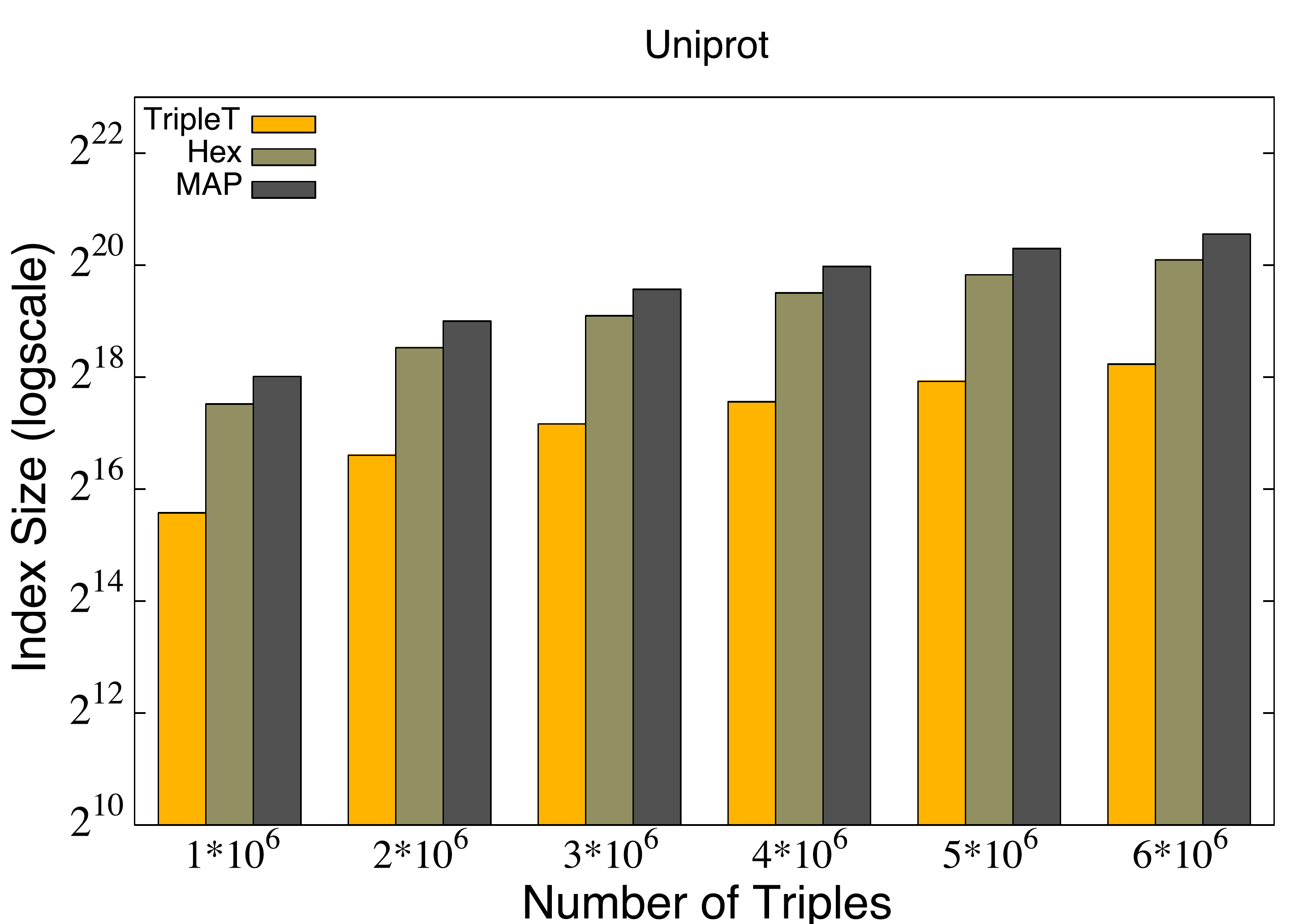}}
\caption{Index sizes, in 8K blocks.}
\end{center}
\end{figure}

In increments of 1 million triples, from 1 to 6 million triples, we built the
three index types.  The plots of the index sizes, in 8K blocks, are shown in
Figures \ref{fig:sizesynth}-\ref{fig:sizeuni}.  TripleT was
up to eight orders of magnitude smaller, with a typical two
orders of magnitude savings in storage cost.  The reason for this
can be attributed to (1) TripleT uses just one B+tree, whereas MAP and HexTree
both require three B+trees, and (2) the key size in TripleT is 1/3 that of MAP and
1/2 that of HexTree, leading to significantly higher branching factor of the
B+tree (and hence shallower trees). 

\subsection{Query performance}

\begin{figure}[!t]
\begin{center}
\subfigure[Synthetic, $k=0$]{\label{fig:k0synth}
\includegraphics[width=0.33\textwidth]{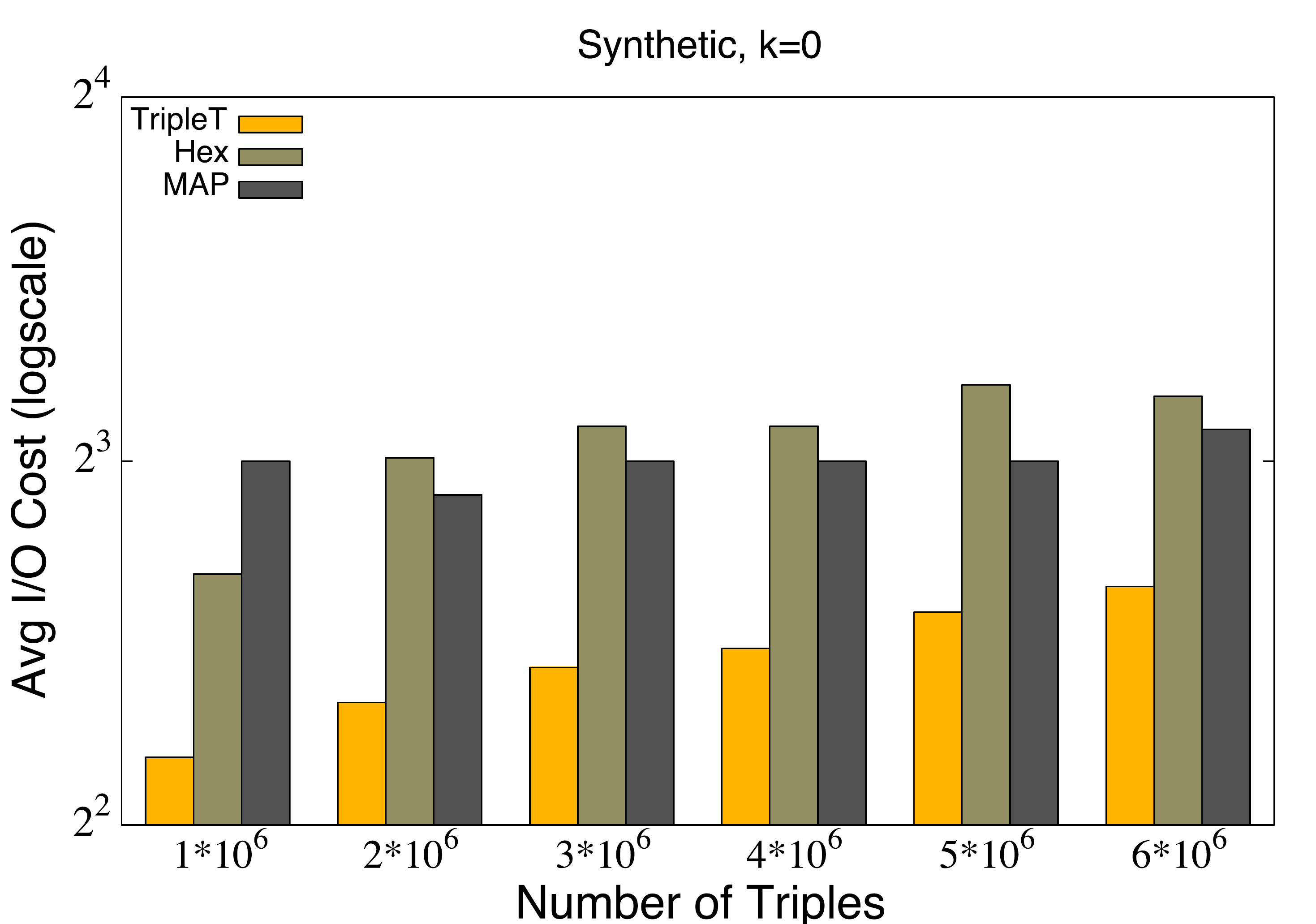}}
\hspace{-.5cm}
\subfigure[DBPedia, $k=0$]{\label{fig:k0db}
\includegraphics[width=0.33\textwidth]{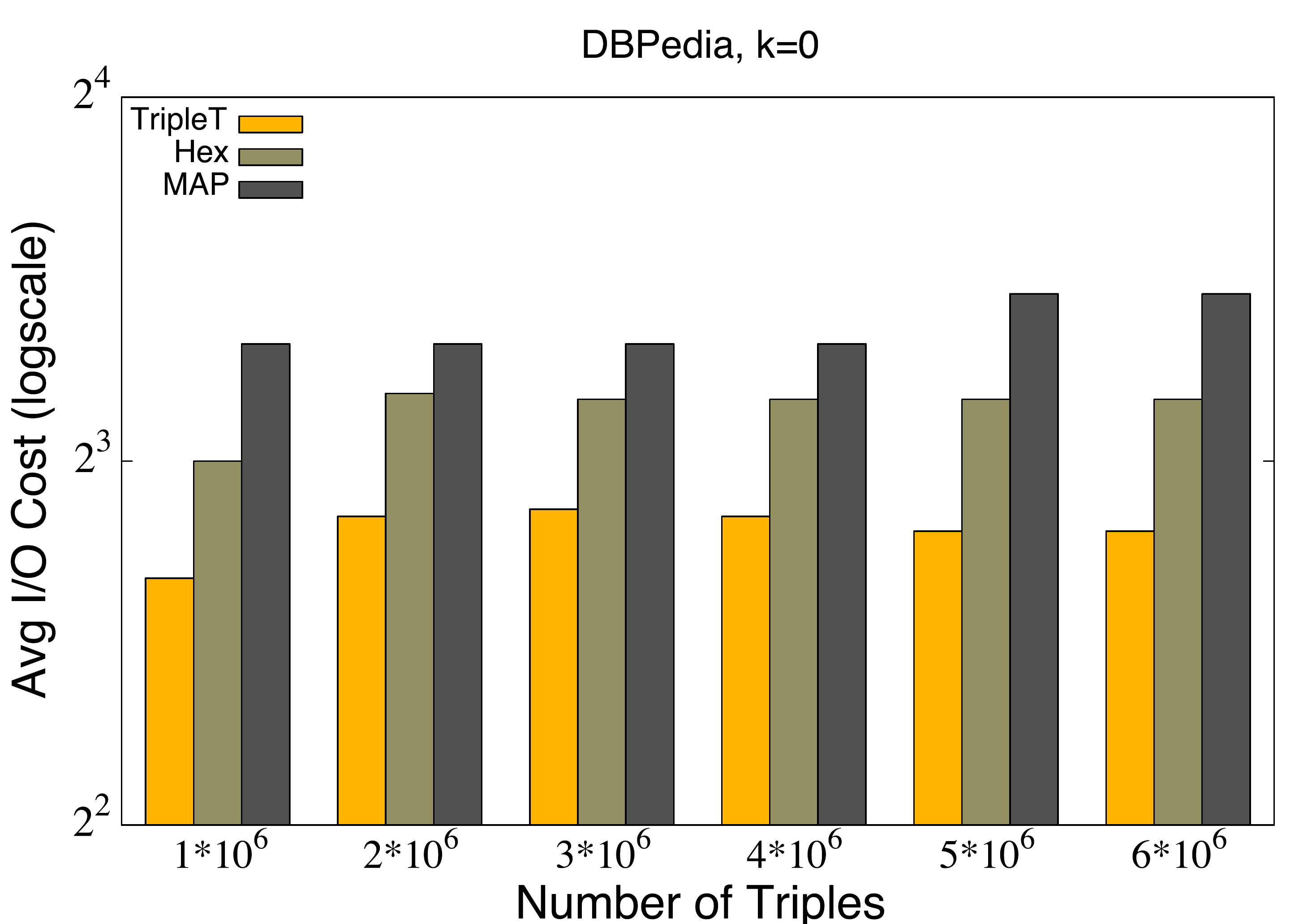}}
\hspace{-.5cm}
\subfigure[Uniprot, $k=0$]{\label{fig:k0uni}
\includegraphics[width=0.33\textwidth]{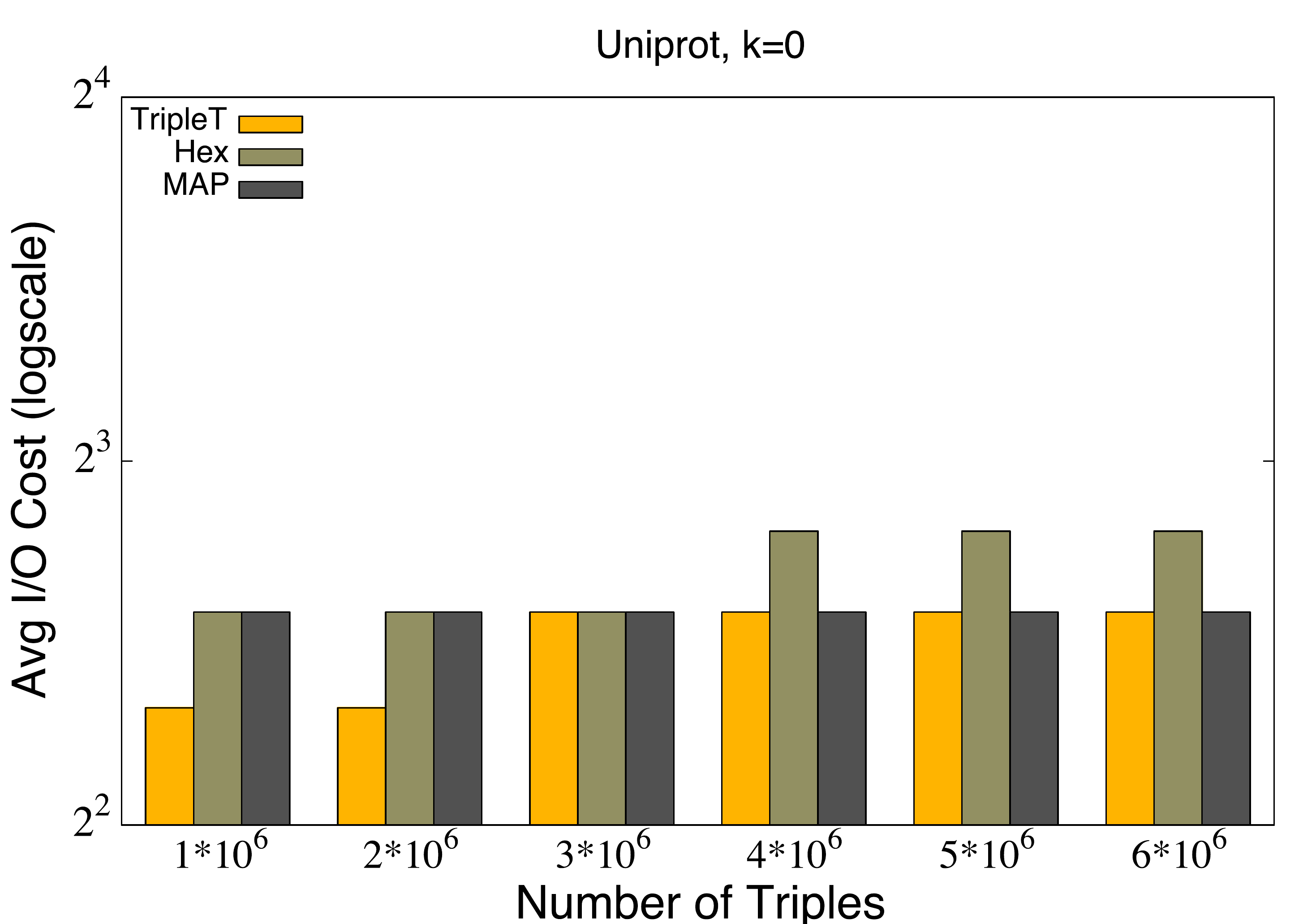}}

~
\vspace{0.3cm}
~

\subfigure[Synthetic, $k=1$]{\label{fig:k1synth}
\includegraphics[width=0.33\textwidth]{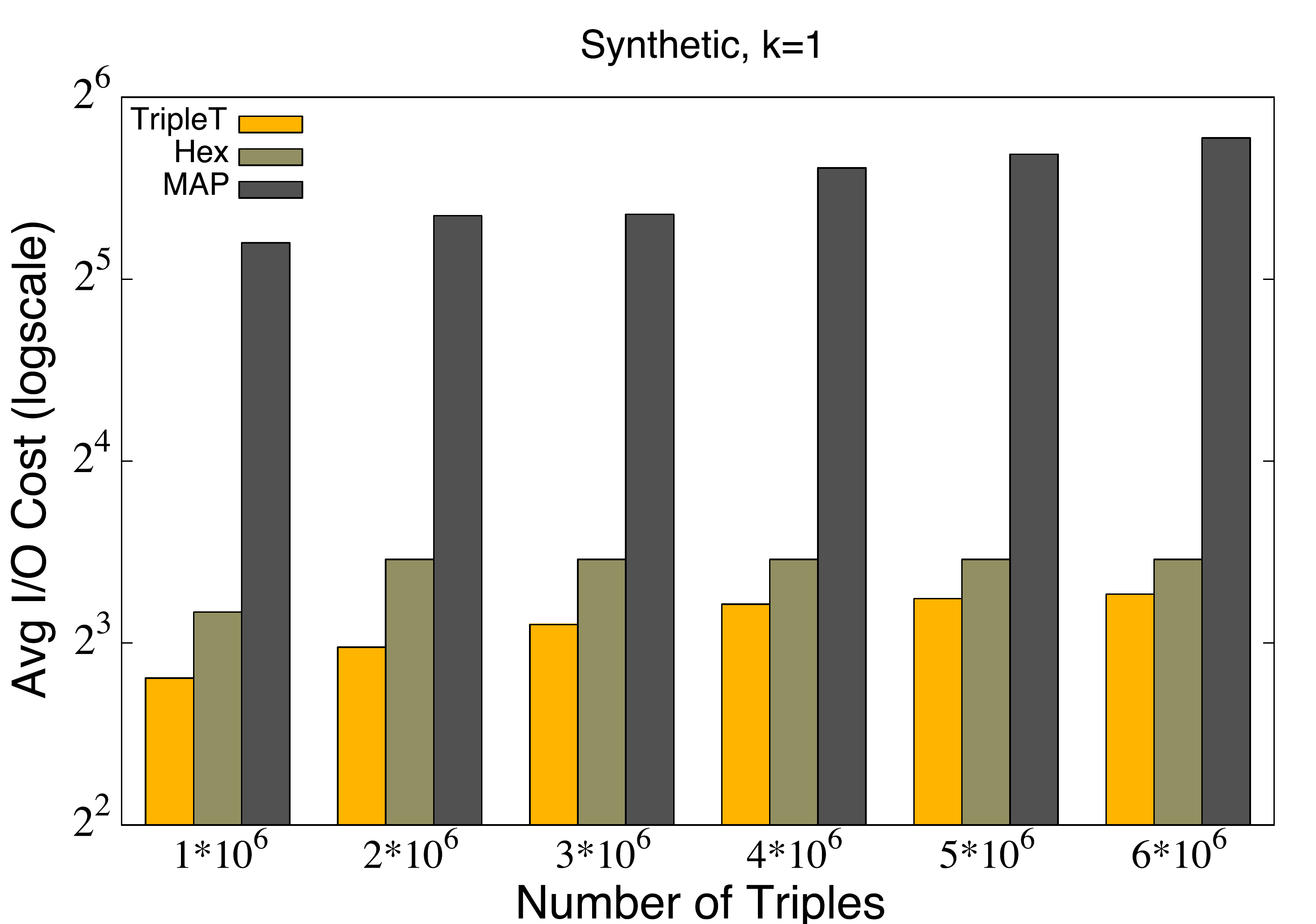}}
\hspace{-.5cm}
\subfigure[DBPedia, $k=1$]{\label{fig:k1db}
\includegraphics[width=0.33\textwidth]{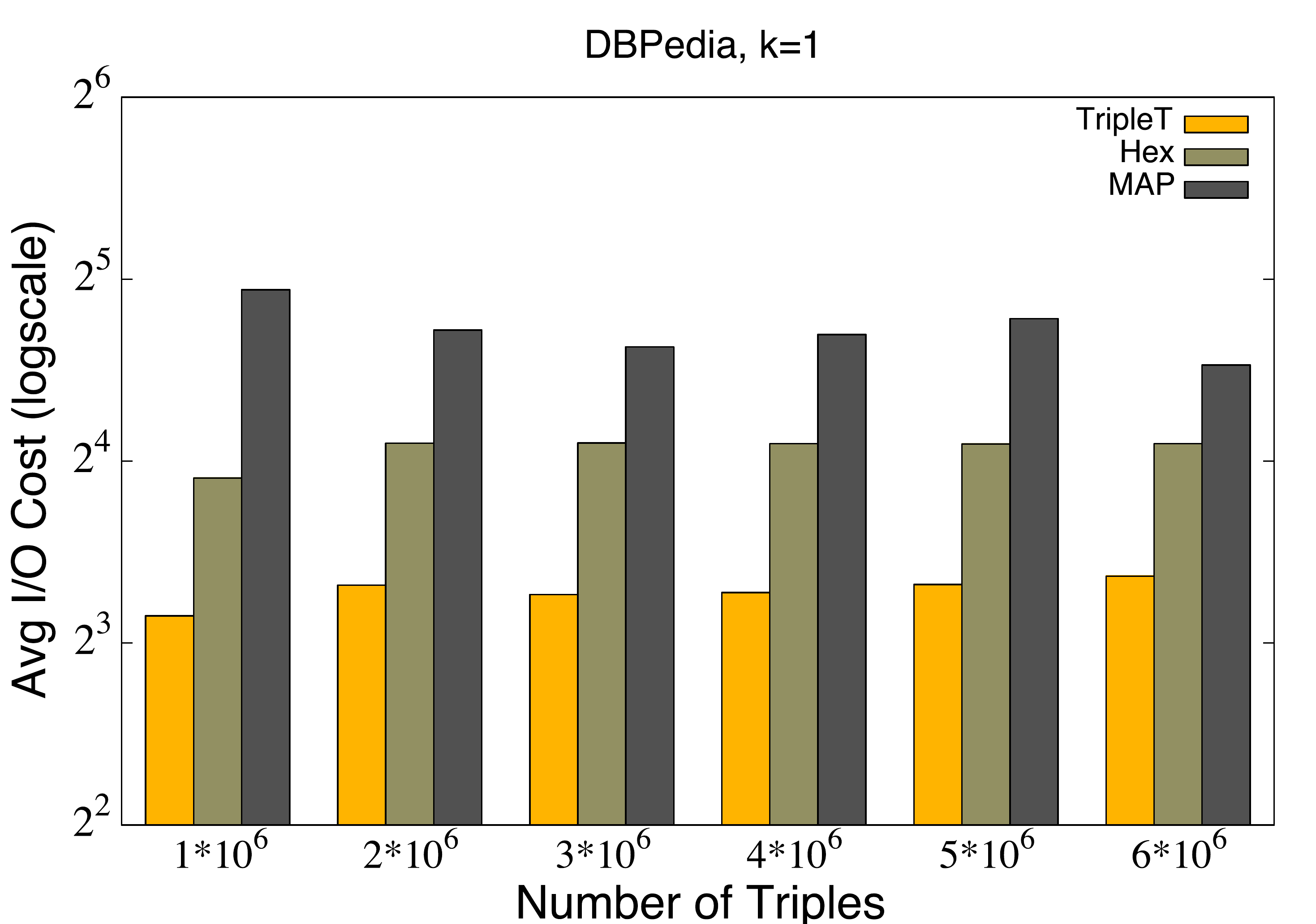}}
\hspace{-.5cm}
\subfigure[Uniprot, $k=1$]{\label{fig:k1uni}
\includegraphics[width=0.33\textwidth]{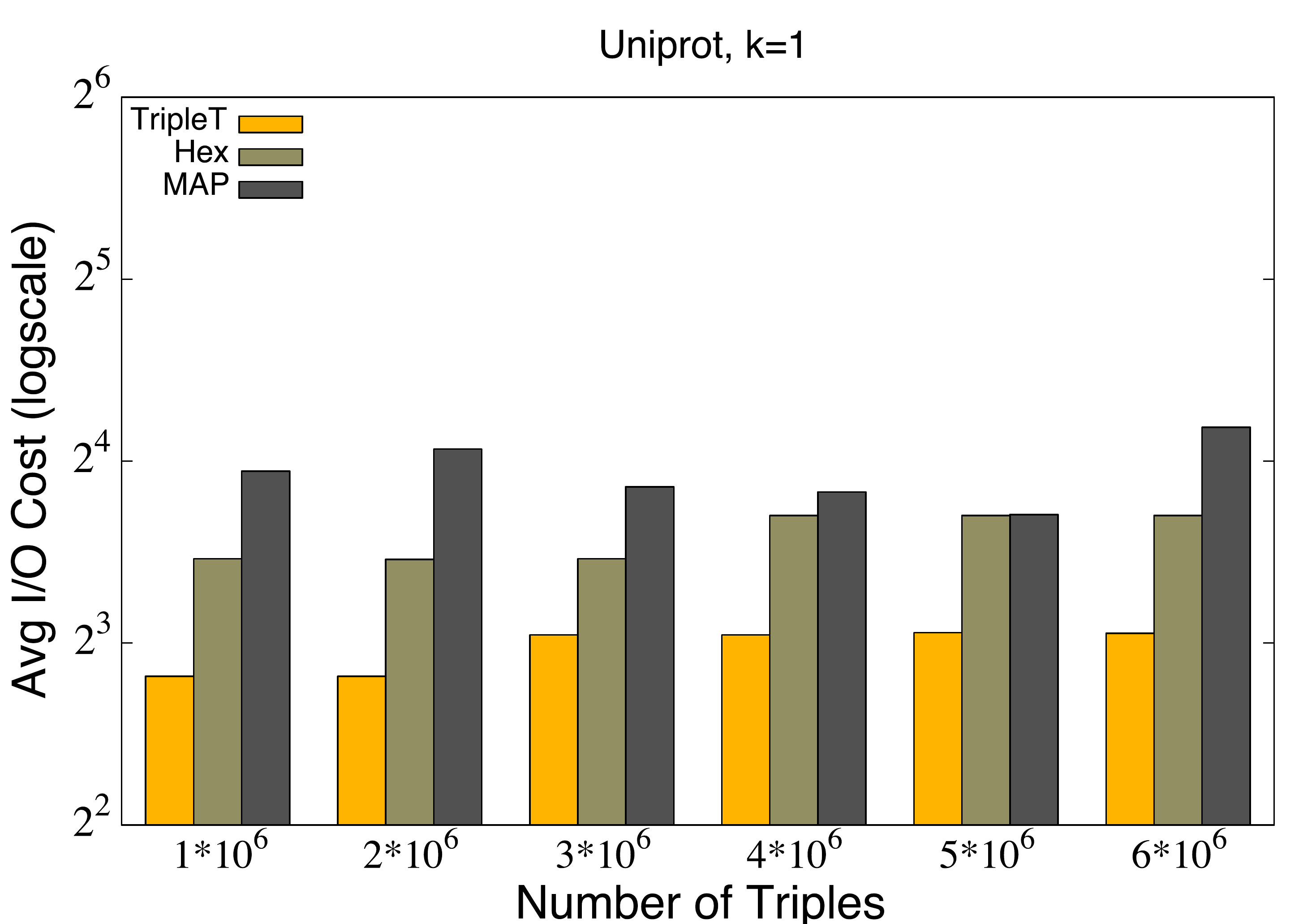}}
\caption{Cost of Query Processing.}
\end{center}
\end{figure}

We use the classic I/O cost model for query evaluation, i.e., we use the
number of block reads as our performance metric \cite{RG},  as we are
interested in comparing the technology-independent behavior of MAP, HexTree, and
TripleT.  We considered two query scenarios:

\begin{itemize}

    \item A single SAP without variables, which we denote as a ``$k=0$'' join
    scenario.  For each dataset, and for each size, we randomly selected ten
    triples from the dataset and recorded the costs of looking them up in MAP,
    HexTree, and TripleT.  The average I/O cost of performing these lookups is
    given in Figures \ref{fig:k0synth}-\ref{fig:k0uni}.  

    \item Basic BGP join patterns, which we denote as a ``$k=1$'' join scenario.  
    We considered four sub-scenarios, covering the basic ways in which SAPs 
    may be joined.
    \begin{enumerate}
        \item Computing the join of two variable-free SAPs having one atom in common.
        \item Computing the join of two SAPs having one atom in common, one SAP
        having a single variable and the other variable-free.
        \item Computing the join of two SAPs having no atoms in common, 
        each having a single variable, which they share.
        \item Computing the join of two SAPs having one atom in common, 
        each having one variable, which they also share.
    \end{enumerate}
    For each data set, for each size, 
    we generated ten random BGPs of each of these four scenarios
    and recorded the cost of their evaluation using MAP, HexTree, and TripleT.  The
    average I/O costs are given in Figures \ref{fig:k1synth}-\ref{fig:k1uni}.  

\end{itemize}

We observe from these experiments that (1) for $k=0$ TripleT never performed
worse than MAP or HexTree, and usually better; and, (2) for $k=1$, TripleT
always out-performed MAP and HexTree, with up to two orders of magnitude improvement
in I/O costs.

\section{Concluding remarks}

It is clear from this extensive evaluation of the full range of BGP join
scenarios on both synthetic and real-world data sets that TripleT is a serious
contender for indexing massive RDF data stores in secondary memory.  Our
proposal is conceptually quite simple, and hence straight forward to implement.
Furthermore, TripleT exhibits multiple orders of magnitude improvement over the
state of the art  for both storage cost and query evaluation cost. In
closing, we note that the many optimizations (such as various key compression
schemes) which have been used in implementations of MAP and HexTree reported
in the literature can
equally be applied to TripleT.

\bibliographystyle{plain}
\bibliography{BGPbib}

\section*{Appendix}

In this section we provide details of the data sets used in the experiments 
discussed in Section \ref{sec:eval}: (1) synthetic data,
(2) the DBPedia RDF data set;\footnote{{\small{\tt http://wiki.dbpedia.org}}} 
and (3) the Uniprot RDF data 
set.\footnote{{\small{\tt http://dev.isb-sib.ch/projects/uniprot-rdf}}}

For (1), we built two synthetic data sets of size 6 million (the results of
Section \ref{sec:eval} are the averages over these two sets). 
In the first set, we randomly generated $n$ triples over $n^{1/3}$ unique atoms, for
$n=1,000,000$, to $n=6,000,000$, in increments of one million, where repetitions
of atoms were allowed within triples.
In the second set, we randomly generated $n$ triples over $ceiling(n^{1/3}) + 2$ unique atoms, for
$n=1,000,000$, to $n=6,000,000$, in increments of one million, where repetitions
of atoms within triples were disallowed.

For (2) and (3),  
we took an arbitrary sample of 10,000,000 triples from each data collection
(treating the DBPedia infobox and pagelinks as one collection) ---
see Table \ref{table:data}.
After cleaning and
duplicate elimination, we kept 6,000,000 triples in each collection.  In this
cleaned data, we use only the first 400 (DBPedia) or 150 (Uniprot) characters of
atoms (note that these are the basis of the fixed key sizes for the B+trees we built).  
This truncation only affected a few extremely long atoms appearing
exclusively in the object position.
Final statistics for these data sets are given in Table \ref{table:stats}.

\begin{table}[!ht]
\begin{center}
\begin{tabular}{c@{~~~~~~~}cc}
$G$ & $|G|$ & average atom length\\
\hline
DBPedia & 82,701,339 & 34.2\\ 
Uniprot & 956,915,180 & 29.0
\end{tabular}
\end{center}
\caption{Data sets}\label{table:data}
\end{table}

\begin{table}[!ht]
\begin{center}
{\footnotesize
\begin{tabular}{cccccccc}
$G$ & $|\subG|$ & $|\predG|$  & $|\objG|$  & $|\allG|$  & $|\subG\cap\objG|$  & $|\subG\cap\predG|$ & $|\predG\cap\objG|$ \\
\hline
DBPedia & 1,370,679 & 20,873 & 1,848,114 & 2,852,484 & 387,182 & 0 & 0\\
Uniprot & 4,357,005 & 81 & 1,734,176 & 5,644,939 & 446,311 & 0 & 12 
\end{tabular}
}
\end{center}
\caption{Basic statistics of sampled data sets}\label{table:stats}
\end{table}

\end{document}